\documentstyle[12pt]{article}

\def\be{\begin{equation}}
\def\ee{\end{equation}}
\def\ba{\begin{array}{c}}
\def\ea{\end{array}}

\def\ben{$$}
\def\een{$$}
\begin{document}

\titlepage
\vspace*{4cm}

\begin{center}{\Large \bf
${\cal PT}-$symmetric harmonic oscillators
 }\end{center}

\vspace{5mm}

\begin{center}
Miloslav Znojil
\vspace{3mm}

\'{U}stav jadern\'e fyziky AV \v{C}R, 250 68 \v{R}e\v{z},
Czech Republic\\

\end{center}

\vspace{5mm}

\section*{Abstract}

Within the framework of the recently proposed formalism using
non-hermitean Hamiltonians constrained merely by their ${\cal PT}$
invariance we describe a new exactly solvable family of the
harmonic-oscillator-like potentials with non-equidistant spectrum.

\vspace{9mm}

\noindent
 PACS 03.65.Ge,
03.65.Fd


\begin{center}
\end{center}

\newpage

\section{Introduction}

An increase of interest in the ${\cal PT}$ symmetric Hamiltonians
\cite{Bessis} - \cite{last} may be explained by a number of their
appealing properties. For illustration, let us pick up the most
ordinary harmonic-oscillator Schr\"{o}dinger equation in one
dimension
 \ben
\left (-\,\frac{d^2}{dr^2} + r^2 \right )
 \, \psi(r) =
E  \, \psi(r), \ \ \ \ \ \psi(r) \in L_2(-\infty,\infty)
 \een
and change its variable $r$ to $x=r+ic$, $c > 0$ in such a way
that $x$ becomes treated as real, $ x \in (-\infty,\infty)$.
Obviously, as long as $r^2=x^2-2icx-c^2$, the asymptotic growth or
decrease of the old general solution $\psi(r)$ remains equivalent
to the asymptotic growth or decrease of the new, complex function
$\varphi(x) \equiv \psi(x-ic)$. In the new, shifted bound-state
problem
 \ben
\left (-\,\frac{d^2}{dx^2} + x^2 -2ic\,x \right )
 \, \varphi(x) =
(E+c^2)  \, \varphi(x), \ \ \ \ \ \varphi(x) \in
L_2(-\infty,\infty)
 \een
boundary conditions remain unchanged, therefore. As a consequence,
the spectrum $E=E_m = 2m+1,  \ m = 0, 1, \ldots $ of energies
remains discrete, real and bounded from below, shifted merely by a
constant $c^2>0$ in the latter case.

Our illustrative non-hermitean Hamiltonian commutes with the
product of parity ${\cal P}$ (changing $x$ to $-x$) and time
reversal ${\cal T}$ (changing, formally, the imaginary unit $i$ to
$-i$). Daniel Bessis \cite{Bessis} and Carl Bender with Stefan
Boettcher \cite{BB} conjectured that such a ${\cal PT}$ symmetry
might be responsible for the reality of spectra for a much broader
class of Hamiltonians. Their conjecture is widely supported by a
number of tests. For the anharmonic $V(x)=x^2+igx^3$, its
perturbative confirmation (viz., the proof using the Borel
summability) has even been made available in the early eighties
\cite{Caliceti}. Very recently, the further numerical and
quasi-classical evidence has been provided by the non-polynomial
$V(x)=x^2(ix)^\delta$ \cite{BB} and by its supersymmetric partners
\cite{Junker} as well as by certain hyperbolic \cite{druhej} and
trigonometric \cite{band} models. An additional, purely
non-numerical backing of the hypothesis may, last but not least,
rely upon the exactly solvable ${\cal PT}$ symmetric $V(x)=\exp(
ix)$ \cite{Junker} and upon the quasi-exactly solvable polynomial
$V(x)=-x^4+iax^3+bx^2+icx$ \cite{BBjpa} and non-polynomial
$V(x)=x^2+iax+(b+icx)/(1+dx^2)$ with $d >0$ \cite{japane}. In the
light of these and further references \cite{last} it is rather
surprising that no attention has been paid, up to now, to harmonic
oscillators with a properly regularized centrifugal-like core of
strength $G = \alpha^2-1/4$ with $\alpha > 0$,
 \be \left (-\,\frac{d^2}{dx^2} + x^2 -2ic\,x
+ \frac{G}{(x-ic)^2} \right )
 \, \varphi(x) =
(E+c^2)  \, \varphi(x), \ \ \ \ \ \varphi(x) \in
L_2(-\infty,\infty).
 \label{SE}
 \ee
The gap is to be filled by the present note. We are persuaded that
this ``forgotten" ${\cal PT}$ symmetric model deserves a few
explicit comments at least.

\section{Quasi-parity}

In the usual one-dimensional and hermitean world the quadratic
singularity proves too strong and, at nonzero $G>0$, it cuts the
real axis in the two separate, non-communicating halves
\cite{Simon}. In the present more permissive context, the threat
is easily avoided by a shift of the singularity off the
integration path. In the complex plane of $x$ cut, say, from
$x=ic$ to $x\to +i\infty$, our problem (\ref{SE}) remains well
defined.

What is equally important is its exact solvability in terms of the
confluent hypergeometric special functions,
 \ben
\varphi(x) =  C_+\,(x-ic)^{-\alpha+1/2}e^{-(x-ic)^2/2}\ _1F_1
\left (  (2-2\alpha-E)/4, 1-\alpha; (x-ic)^2
 \right )
+
 \een
 \ben
+
 C_-\,(x-ic)^{\alpha+1/2}e^{-(x-ic)^2/2}\ _1F_1 \left (
 (2+2\alpha-E)/4, 1+\alpha; (x-ic)^2
 \right ).
 \een
For large $|x|$ this expression grows as $\exp(x^2/2)$ and
violates boundary conditions unless it degenerates to a polynomial
(cf., {\it mutatis mutandis}, \cite{Fluegge}). In this way we get
the (complete) spectrum of energies
 \ben
 E=E_{qn}=4n+2 + 2 q \alpha
 \een
numbered by the quasi-parity $q = \pm $ and integers  $ n = 0, 1,
2, \ldots$. The related normalizable wave functions
  \be
\varphi(x) = const. \,(x-ic)^{-q \alpha+1/2}e^{-(x-ic)^2/2} \
L^{(-q \alpha)}_n \left [
 (x-ic)^2
 \right ]
\label{waves}
 \ee
are defined in terms of the well known orthogonal Laguerre
polynomials,
 \ben
 \ba
 L^{\beta}_0(z)=1,\\
 L^{\beta}_1(z)=\beta+1-z,\\
 L^{\beta}_2(z)=(\beta+2-z)^2 -(\beta+2),\\
 L^{\beta}_3(z)=(\beta+3-z)^3 -3(\beta+3)(\beta+3-z)
 +2(\beta+3),\\
 \ldots .
  \ea
 \een
In the limit $ \alpha \to 1/2$ and $c \to 0$ our Hamiltonian
re-acquires its hermiticity. Our set of solutions coincides with
the well known one-dimensional harmonic oscillators and the
quasi-parity degenerates to the ordinary parity,
 \ben
{\cal P} \psi(r) = \psi(-r) = q \,\psi(r).
 \een
The spectrum of energies becomes equidistant, $E_{+0}=1$,
$E_{-0}=3$, $E_{+1}=5$, $E_{-1}=7$ etc. Precisely $2n+(1-q1)/2$
real nodal zeros appear in the corresponding real wave functions.

After we switch on a ``subcritical", permitted central attraction
on, the nodal zeros of $\varphi(x)$ in eq. (\ref{waves}) will move
upwards in the complex plane. Within interval $-1/4<G<0$ with
$0<\alpha <1/2$, all the even and odd energies undergo an upward
and downward constant shift, respectively. At the infinitesimally
small extreme values of $\alpha \approx 0$ all the energies almost
degenerate in doublets $E_{\pm 0} \approx 2$, $E_{\pm 1}\approx
6$, $E_{\pm 2}\approx 10$ etc.

\section{Strong repulsion and level crossing}

We have seen that the ${\cal PT}$ symmetry acts (or at least might
act) as a simple and efficient means of regularization of a
singularity in $V(x)$ in one dimension \cite{japane}. Our present
example with $\alpha>1/2$ extends this idea in a way inspired by
the solvability of the radial equation in three dimensions. There,
$G=\ell(\ell+1)$ contains the angular momentum $\ell=0,1,\ldots$
and may become quite large. After the present regularization and
analytic continuation of this model to one dimension the only
important novelty is the sudden disappearance of the (now,
redundant) boundary condition in the origin.

In the language using the complex coordinates $x \in C\!\!\!\!I$
we may also return from one dimension with $x\in(-\infty ,\infty)$
to three or more dimensions with $x \in (0, \infty)$. Beyond such
a purely kinetic interpretation of our strongly repulsive core a
genuine dynamical meaning of  $G= \ell(\ell+1)\gg 1$ may be
encountered, say, in nuclear physics where one has to use $\ell =
3, 33, 117, 352, 517$ and $ 1083$ in an efficient approximative
description of the respective nuclei $^4He$, $^{16}O$, $^{40}Ca$,
$^{90}Zr$, $^{120}Sn$ and $^{208}Pb$ in the so called breathing
mode \cite{Sotona}.

After we turn on an enhanced repulsion in eq. (\ref{SE}) we
discover a quasi-degeneracy and crossing of levels with opposite
quasi-parities in the vicinity of every integer $\alpha = 1, 2,
\ldots$. In the very first case with $\alpha=1$ we may factor the
square $(x-ic)^2$ out of the states with the even quasi-parity,
 \ben
 \ba
 L^{(-1)}_0\left [ (x-ic)^2 \right ]=1,\\
 L^{(-1)}_1\left [ (x-ic)^2 \right ]=-(x-ic)^2 ,\\
 L^{(-1)}_2\left [ (x-ic)^2 \right ]=-2(x-ic)^2+ (x-ic)^4,\\
 L^{(-1)}_3\left [ (x-ic)^2 \right ]=-(x-ic)^2 \left [
(x-ic)^4-6 (x-ic)^2+6 \right ]
 \\
 \ldots\ .
  \ea
 \een
The resulting formula $ L^{(-1)}_{n+1}\left [ (x-ic)^2 \right
]=-(x-ic)^2\,L^{(1)}_n\left [ (x-ic)^2 \right ]$ implies that the
even and odd quasi-parity partners will coincide {\em precisely}
at the ``exceptional" \cite{Simon} value of $G=3/4$. An unavoided
crossing of the energy levels occurs {\em without} their
degeneracy. Similar phenomenon may be observed at all the
subsequent integers $\alpha = 2, 3, \ldots\ $.

During the steady growth of the repulsion $\alpha>1/2$ the
relative displacement of the two halves of the spectrum
distinguished by their quasi-parity is accompanied by certain
interesting changes in the structure and position of the nodal
zeros in $\varphi(x)$. Their detailed analysis already lies out of
the scope of the present note. Mathematically, it reflects a
complex generalization of the usual Sturm-Liouville oscillation
theorems \cite{Hille}.

\section{Weak core as a perturbation}

Besides the natural interpretation of small deviations from
equidistant spectrum in a weak-coupling regime with $G \approx 0$
we may also try to trim or suppress the influence of the core in
eq. (\ref{SE}) via a sufficiently large screening $c \gg 1 $. In
such an alternative setting our potential may be decomposed into
its dominant (shifted) harmonic oscillator part $V^{(HO)}(x) -
(x-ic)^2$ and a well-behaved ${\cal O}(1/c^2)$ perturbation,
 \ben
V(x) = V^{(HO)}(x) + G W(x), \ \ \ \ W(x) = W^{(I)}(x)
+W^{(II)}(x)+W^{(III)}(x).
 \een
After a re-parameterization $\mu =g= c^{-2}$ and $\lambda =
-c^{-4}$ the first, asymptotically dominant ${\cal O}(x^{-2})$
component of the anharmonicity
 \ben
 W^{(I)}(x) = \frac{1}{x^2+c^2} \equiv \mu +
\frac{\lambda\,x^2}{1+g\,x^2}
 \een
appears quasi-exactly solvable at certain strengths $G$
\cite{Flessas}. This correction has already been used in numerous
methodical considerations \cite{Gallas}. The subsequent term
 \ben
W^{(II)}(x) = i\frac{2cx}{(x^2+c^2)^2}= {\cal O} (1/x^3)
 \een
is less common. It does not commute with the parity ${\cal P}$ and
breaks the hermiticity of the (unshifted) oscillator, obeying only
the overall ${\cal PT}$ invariance. The last, real and even
component
  \ben
W^{(III)}(x)  = -\frac{2c^2}{(x^2+c^2)^2}= {\cal O} (1/x^4)
 \een
converts the perturbation $W(x)$ to its present exactly solvable
form. It is bounded and asymptotically decreasing. Its routine
treatment, say, within the Rayleigh-Schr\"{o}dinger perturbation
formalism may be expected nicely convergent \cite{Kato}.

\newpage
\section{Summary}

Many potentials of phenomenological interest are analytic
functions. This makes (or at least might make) the underlying
differential Schr\"{o}dinger equation and many properties of its
solutions much more transparent. In particular, we may imagine
that all the functions which satisfy the equation on a real
interval may be immediately continued into a bigger complex
domain.

In principle, the related possible shift or deformation of the
axis of coordinates breaks the hermiticity of the Hamiltonian.
Bound states acquire ${\rm Im}\ E \neq 0$ and become
re-interpreted as unstable resonances \cite{Horacek}. With quite a
few important exceptions: It is already known for many years that
certain non-hermitean Hamiltonians $H \neq H^+$ still do support
perfectly stable bound states. The puzzling existence of these
exceptional ``stable resonances" with ${\rm Im}\ E = 0$ could
prove helpful in phenomenological considerations and has been
subject to an intensive study recently.

In this context, our present note has shown that in the particular
quantization scheme which weakens the hermiticity of a Hamiltonian
to its mere ${\cal PT}$ invariance the one-dimensional
superposition $V(r) = r^2+ G/r^2$ of the harmonic and
centrifugal-like forces may be regularized by a purely imaginary
shift of $r$ in such a way that the whole model remains exactly
solvable.

\end{document}